\documentclass[a4paper,11pt]{article}
\usepackage{subfiles}
\usepackage{pos}
\usepackage{xcolor}
\usepackage{graphicx}
\newlength{\bibitemsep}
\newlength{\bibparskip}
\let\oldthebibliography\thebibliography
\renewcommand\thebibliography[1]{%
  \oldthebibliography{#1}%
  \setlength{\parskip}{\bibitemsep}%
  \setlength{\itemsep}{\bibparskip}%
}

\title{Status and plans for the instrumentation of the IceCube Surface Array Enhancement}

\ShortTitle{Surface Array Enhancement: IceCube}

\author{The IceCube Collaboration \\{\normalsize \normalfont(a complete list of authors can be found at the end of the proceedings)}\\}

\emailAdd{shefali.shefali@kit.edu}
\emailAdd{frank.schroeder@kit.edu}
\abstract{
The surface array of IceCube, IceTop, operates primarily as a cosmic-ray detector, as well as a veto for astrophysical neutrino searches for the IceCube in-ice instrumentation. However, the snow accumulation on top of the IceTop detectors increases the detection threshold and attenuates the measured IceTop signals. Enhancing IceTop by a hybrid array of scintillation detectors and radio antennas will lower the energy threshold for air-shower measurements, provide more efficient veto capabilities, enable more accurate cosmic-ray measurements, and improve the detector calibration by compensating for snow accumulation. After the initial commissioning period, a prototype station at the South Pole has been recording air-shower data and has successfully observed coincident events of both the scintillation detectors and the radio antennas with the IceTop array. The production and calibration of the detectors for the full planned array has been ongoing. Additionally, one station each has been installed at the Pierre Auger Observatory and the Telescope Array for further R\&D of these detectors in different environmental conditions. This contribution will present the status and future plans of the hybrid detector stations for the IceCube Surface Array Enhancement.
\vspace{4mm}
{\bfseries Corresponding authors:}
S. Shefali$^{1*}$, Frank G. Schroeder$^{2,3}$\\
{$^{1}$ \itshape Institute of Experimental Particle Physics, Karlsruhe Institute of Technology (KIT), Germany}\\
{$^{2}$ \itshape Institute for Astroparticle Physics, Karlsruhe Institute of Technology (KIT), Germany}\\
{$^{3}$ \itshape  Bartol Research Institute and Dept. of Physics and Astronomy, University of Delaware, USA}\\ [4mm]
$^*$ Presenter

\ConferenceLogo{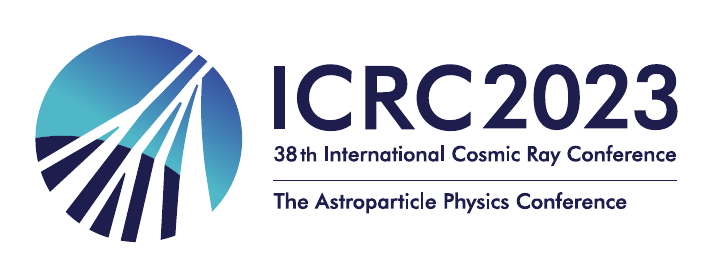}

\FullConference{The 38th International Cosmic Ray Conference (ICRC2023)\\ 26 July -- 3 August, 2023\\ Nagoya, Japan}
}

\begin{document}
\maketitle

\section{Introduction}\label{sec1}

IceTop~\cite{icetop_paper}, the surface array of the IceCube Neutrino Observatory~\cite{Aartsen:2016nxy}, is a unique cosmic-ray detector, which contributes significantly to the veto against the atmospheric background for the In-Ice detector. It consists of 81 pairs of Ice-Cherenkov tanks, placed on the surface of the antarctic ice covering an area of $1~\mathrm{km^2}$. The complementary information from the surface and the In-Ice detector allows for a range of cosmic ray studies, including mass composition, energy spectra, muon density~\cite{soldin2022cosmic}~\cite{Aartsen_2019} etc., in the energy range of 250~TeV to EeV. Since their deployment, non-uniform snow accumulation on the IceTop tanks has resulted in an increase in the detection threshold, due to attenuation in the IceTop signals. To lower this threshold, and develop a multi-component cosmic-ray detection infrastructure, an enhancement of the surface array has been proposed~\cite{Haungs_2019}.

\begin{figure}[h]
  \centering
  \includegraphics[scale=0.39]{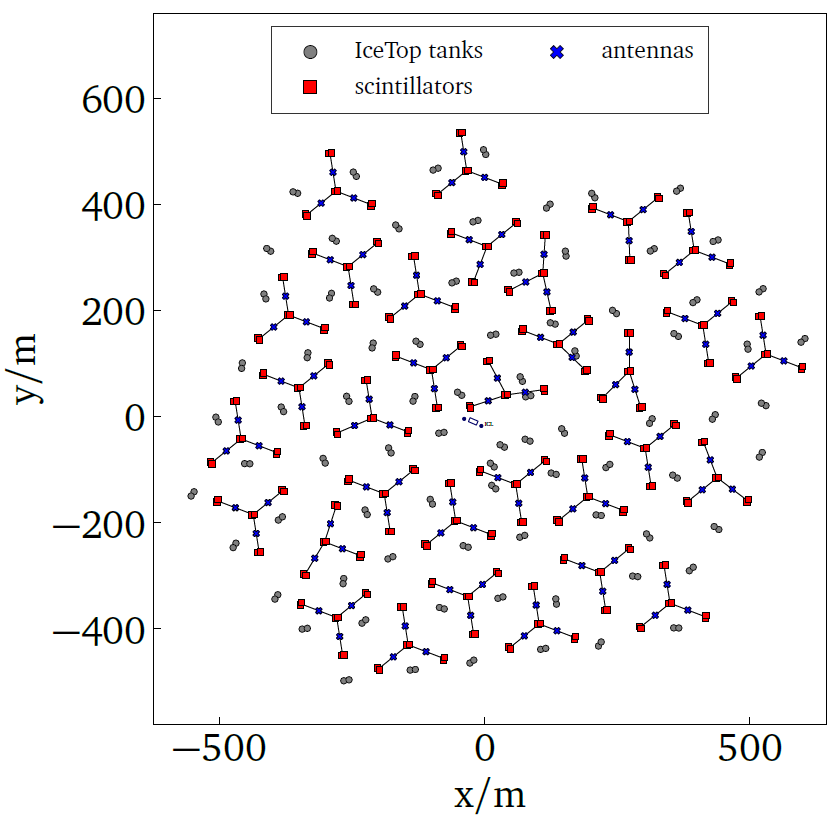}
   \includegraphics[scale=0.39]{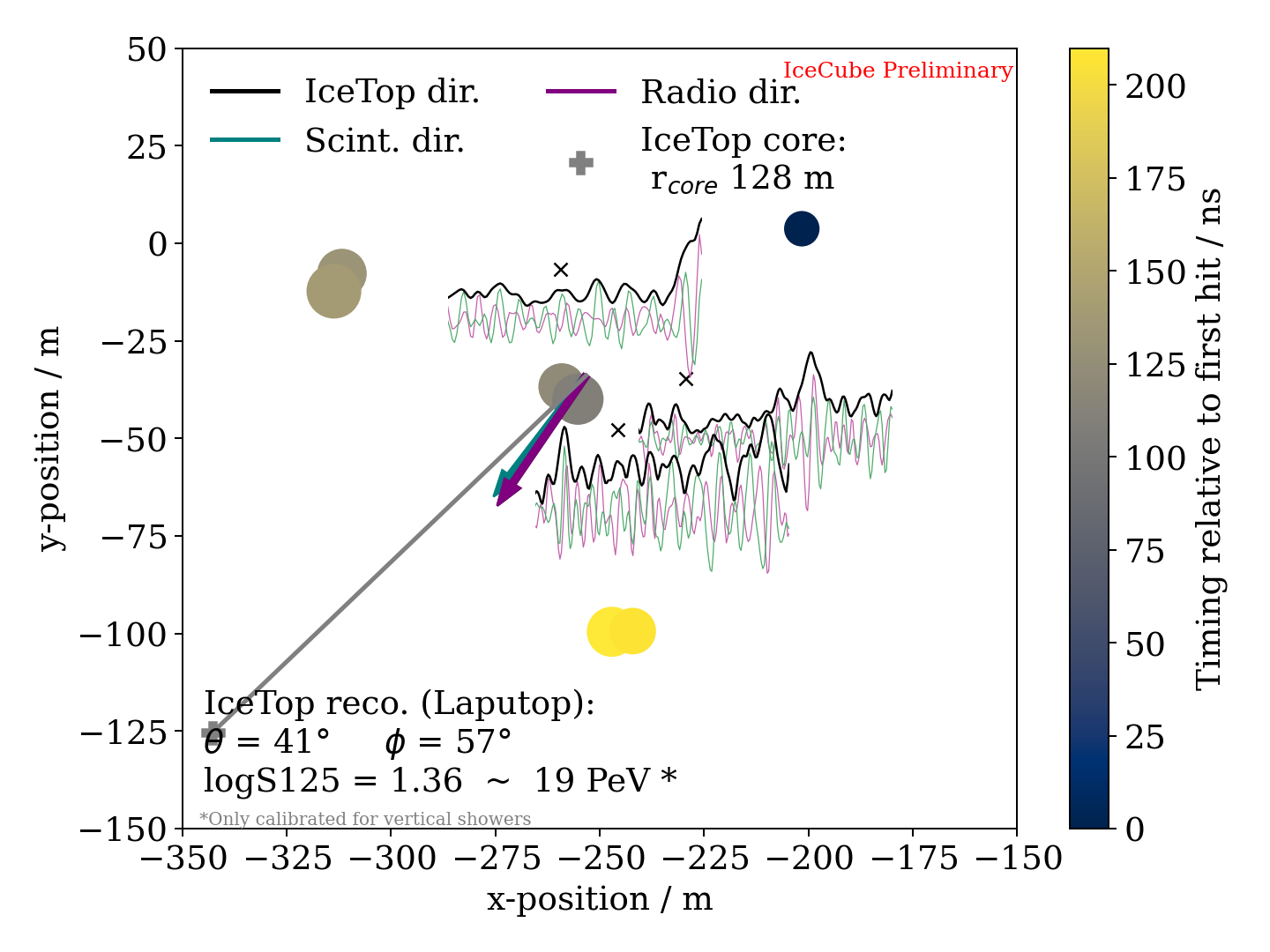}
  \caption{Left: The planned Surface Array Enhancement within the IceTop footprint; Right: An example event with 3-fold coincidence observed by IceTop, scintillation and radio detectors.}
  \label{foot_exevent}
  \vspace*{-0.3cm}
\end{figure}

The enhancement is planned as a hybrid detector array, consisting of 32 stations within the footprint of IceTop (Fig.~\ref{foot_exevent} (left)). Each station will comprise 8 scintillation detectors, 3 radio detectors and a central fieldhub data acquisition (DAQ) system. Following the intermediate R\&D deployments, a fully functional prototype station was operational at the South Pole from January 2020 to December 2022. With the air-shower measurements from the prototype station alone, 3-fold coincident events with the IceTop array were already reconstructed~\cite{dujmovic2021airshower}. An example of such an event is presented in Fig.~\ref{foot_exevent} (right). Events coincidentally observed within $2~\mathrm{\mu s}$ by all 3 detectors are considered to be a single coincident event.  The timestamp of the first scintillation detector hit is taken as the reference time for the coincidence for the scintillator channel, while for the radio detection their trigger time is used. For the IceTop data the timestamp corresponds to when the shower is expected to hit the surface.

Efforts to significantly improve the isolation of the radio emission of the air showers from that of the Galactic and extragalatic background noise, using machine learning methods called convolution neural networks, are ongoing~\cite{AbdulProceeding}. A successful implementation on a 2 month data set from the prototype station has been achieved. The method is able to successfully distinguish and denoise the air-shower signals, and is found to be in agreement with the traditional IceTop reconstruction. Furthermore, the $X_\mathrm{max}$ analysis of the air showers, which can be used to facilitate mass composition studies, using this data set is also in progress. Preliminary results can be found in~\cite{OurProceeding}.

Succeeding the prototype station, the first enhancement station has been deployed at the South Pole in January 2023. Moreover, the production of 1/4th of the planned enhancement stations has been completed. This contribution will present the deployment and commissioning of the new station, as well as the status and plans for the mass production of the proposed stations for future deployments.

\section{Station 0}\label{sec1}

\begin{figure}[h]
  \centering
  \includegraphics[scale=0.28]{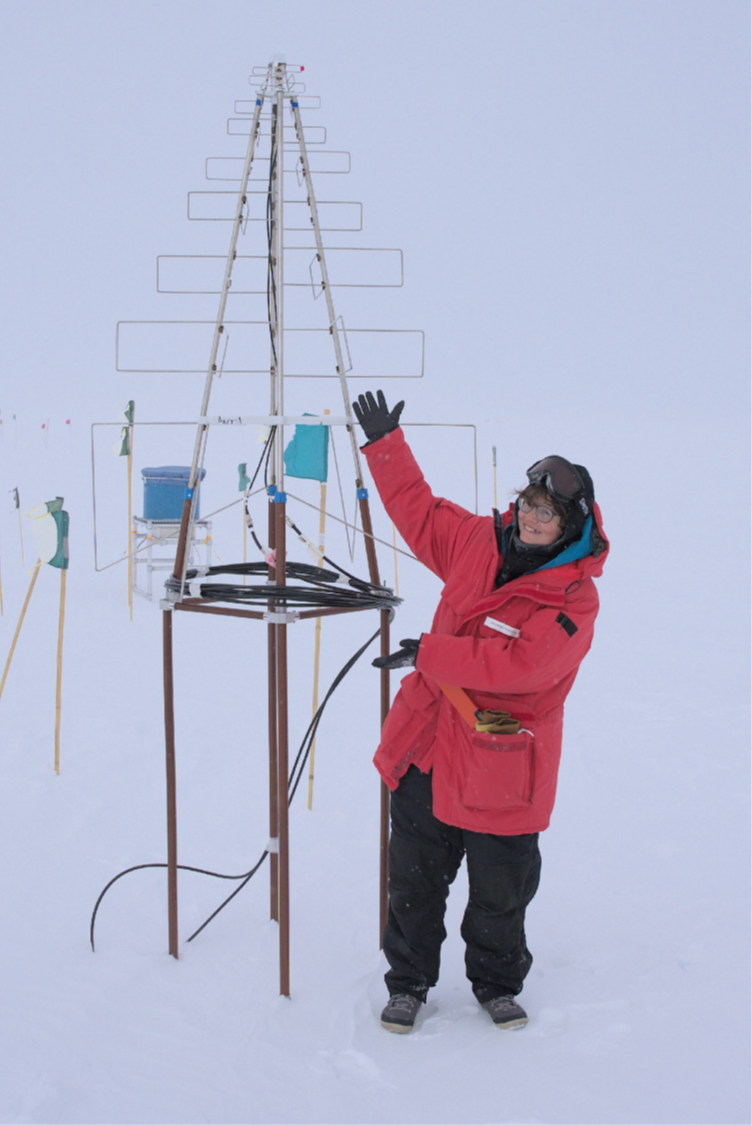}
  \includegraphics[scale=0.1]{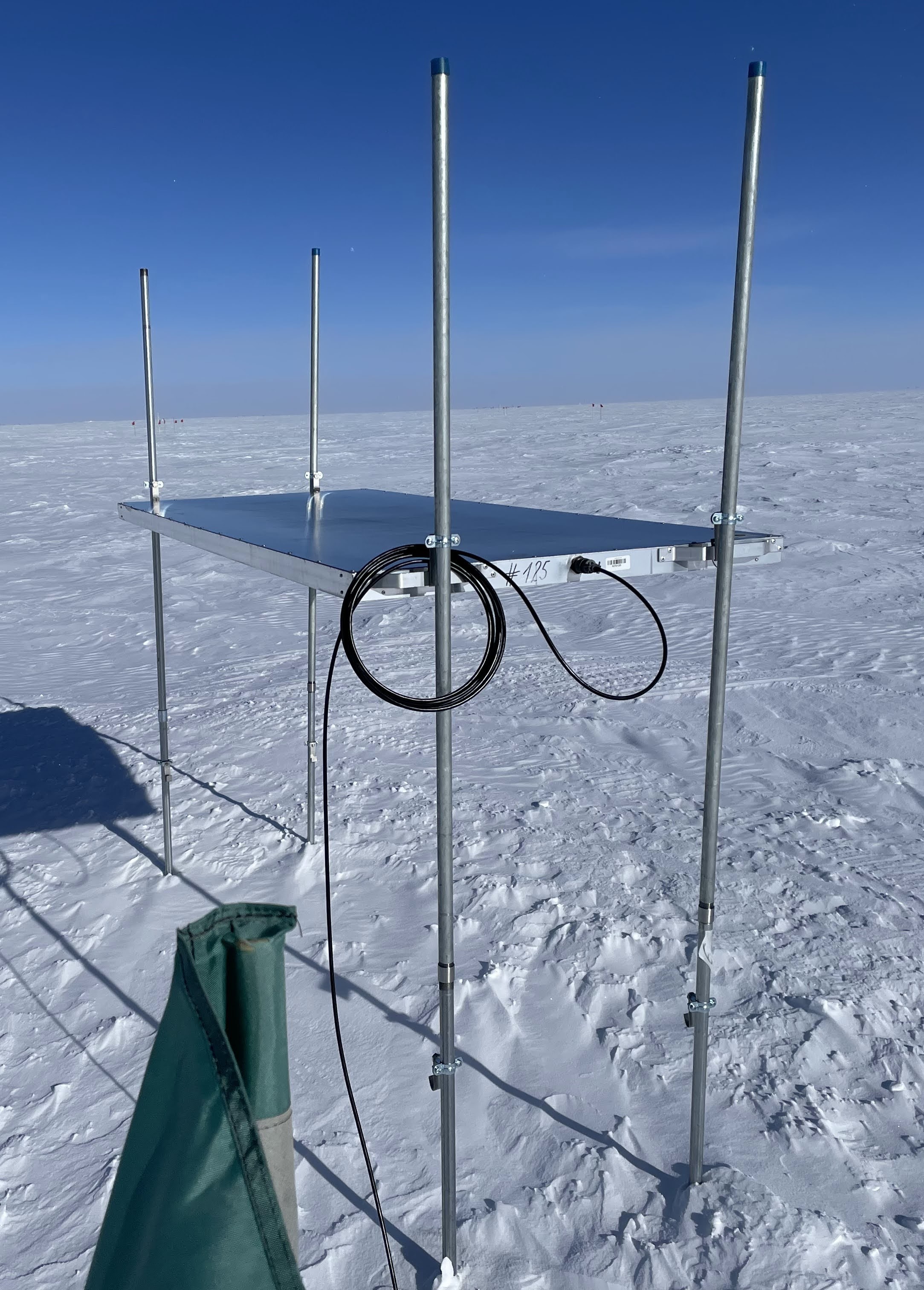}
  \caption{Left: The new deployed antenna with the custom mount; Right: One of the new scintillation detectors.}
  \label{deploy_pics}
  \vspace*{-0.3cm}
\end{figure}

The most recent deployment for the Surface Array Enhancement was in January 2023. This deployment included an exchange of the 8 scintillation detectors of the existing prototype station, and one of the radio antennas. The newly installed scintillation detectors are equipped with an updated version of the readout electronic (uDAQ) board~\cite{oehler2021development}, which was modified to increase the dynamic range of the detectors. Additionally, a mechanical improvement in the detector casing by gluing the edges (in addition to riveting), was introduced to make the panels completely light tight. The new antenna was installed on the custom mount~\cite{roxanne_thesis}, with a pre-calibrated low noise amplifier (LNA). The remaining two antennas were already placed on the mounts in previous deployments. This resulted in the first station of the surface enhancement, \emph{station 0}. Additionally, to avoid the snow accumulation in the coming years, the scintillation detectors, as well as the fieldhub, were raised to $\approx$~1.5~m height. Fig.~\ref{deploy_pics} shows the deployment of the antenna and one of the new panels.

Following the deployment, a commissioning period ensued, which was used for calibration of the updated scintillation detectors. No additional calibration was required for the swapped antenna since the new antenna electronics were pre-calibrated and do not exhibit much variation with temperature~\cite{roxanne_thesis}.

\begin{figure}[t]
  \centering
  \includegraphics[scale=0.45]{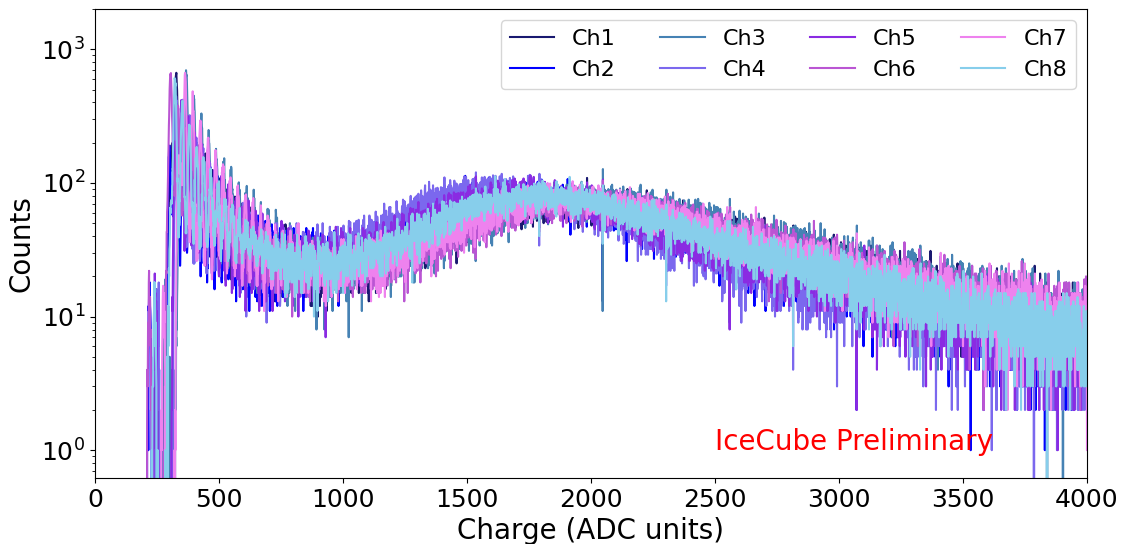}
  \caption{Charge histogram data for the 8 scintillation detectors.}
  \label{multi}
  \vspace*{-0.3cm}
\end{figure}

Fig.~\ref{multi} shows an example charge histogram of a single calibration measurement for all the 8 scintillation detectors, prior to the calibration. Along with proof of functionality, an agreement in behavior is evident for all the scintillation detectors. Owing to this, during the commissioning period, daily air-shower measurement runs were continued, along with calibration runs. For these air-shower measurements, the scintillator threshold was set to the highest digital channel ($\approx$3600 ADC units~\cite{Oehler2022_1000142813}) where the charge deposit is comparable for all detectors, and the multiplicity for sending a trigger to the radio antennas was set to 6 or more triggered scintillation detectors. Whereas, the calibration runs included histogram measurements (in which data is histogrammed on the uDAQ board itself) at a lower threshold with a variation in the supplied bias voltage.

\begin{figure}[h]
  \centering
  \includegraphics[scale=0.51]{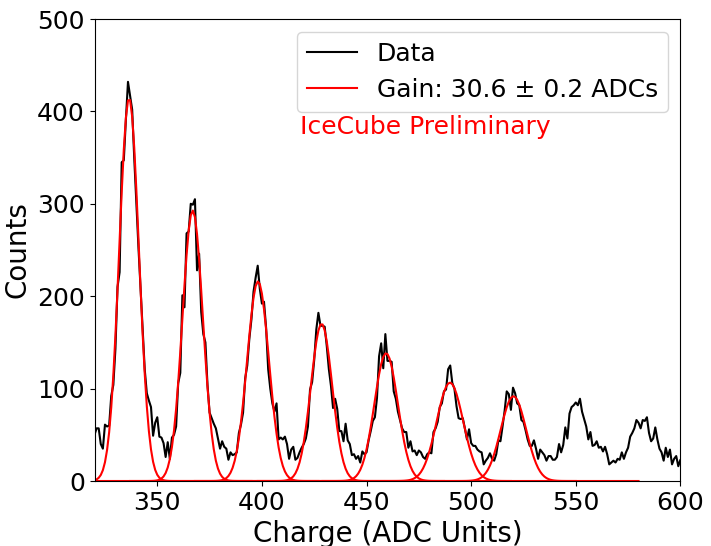}
  \includegraphics[scale=0.32]{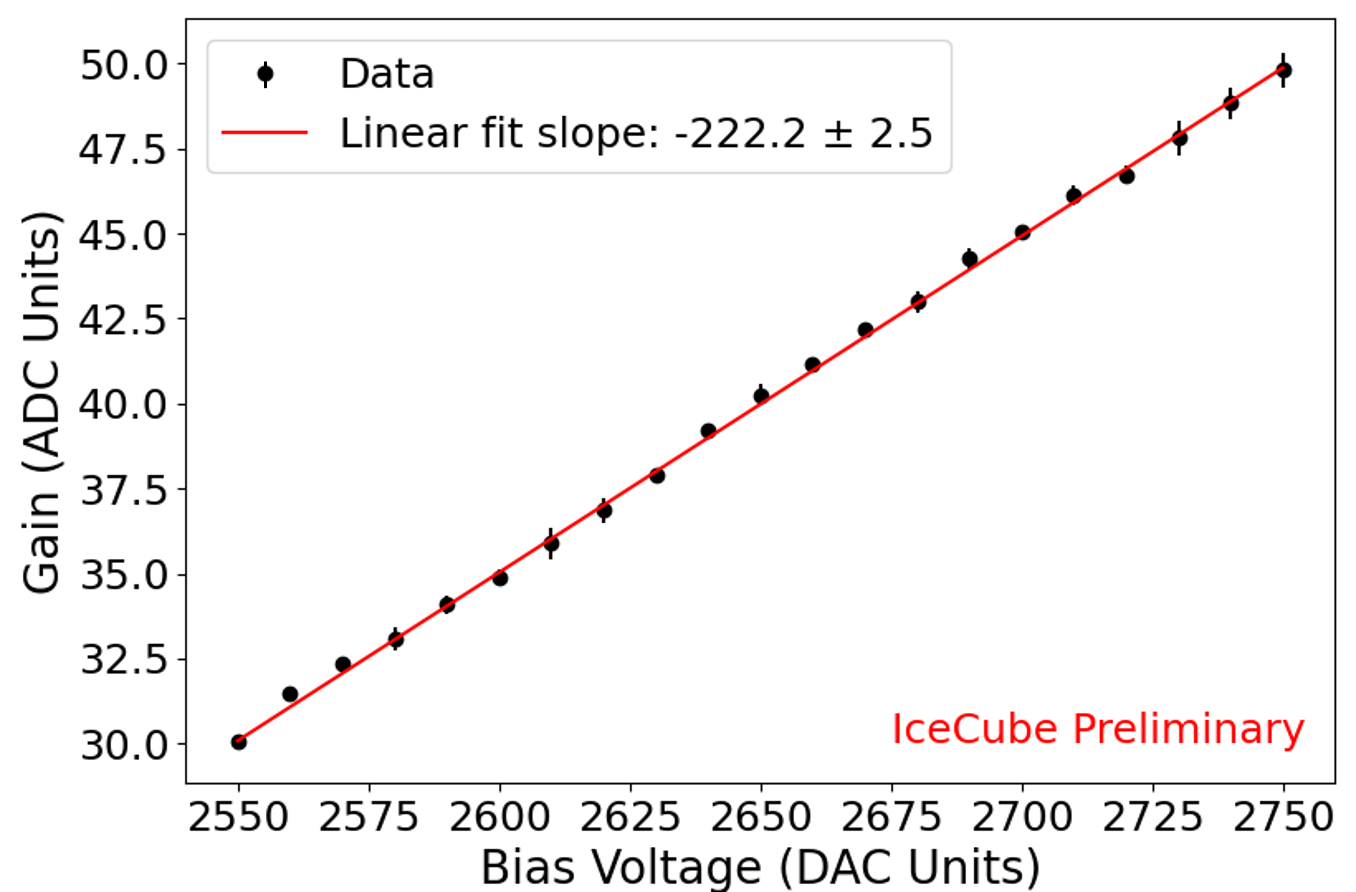}
  \caption{Features of the charge histogram data from the scintillation detectors. Left: Single Photo Electron (SPE) peaks; Right: Gain versus Bias Voltage for a single calibration run.}
  \label{finger_gain}
  \vspace*{-0.3cm}
\end{figure}

\begin{figure}[h]
  \centering
    \includegraphics[scale=0.55]{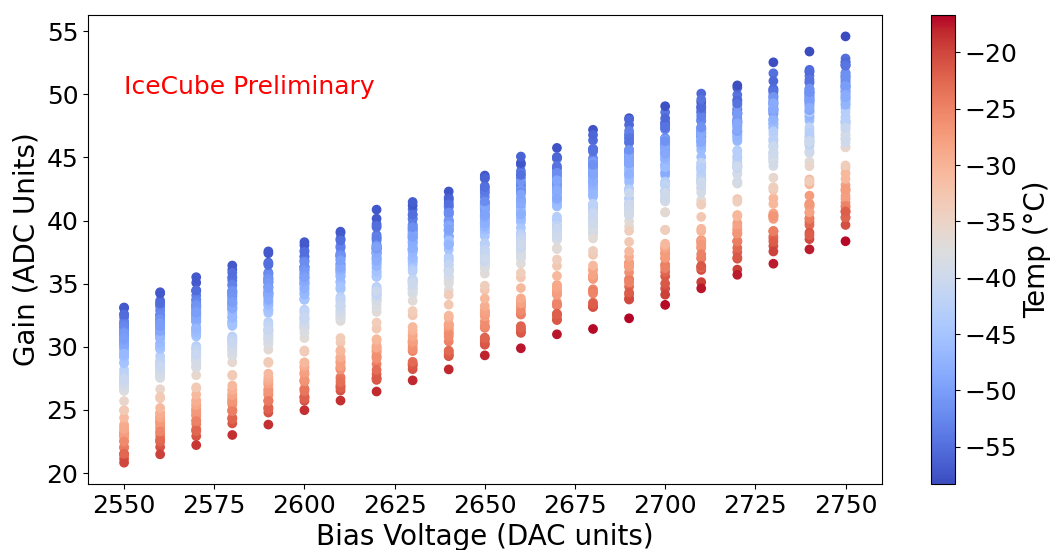}
  \caption{Gain as a function of the temperature and bias voltage from calibration data.}
  \label{3D}
  \vspace*{-0.3cm}
\end{figure}

The charge histograms obtained from the calibration runs were further investigated, and some of the features are shown in Fig.~\ref{finger_gain}. The scintillation light produced in the detectors is recorded by a Silicon Photo Multiplier (SiPM). At the low temperature environment of the South Pole (\nobreakdash-40 to \nobreakdash-70~$^{\circ}$C), the gain of these SiPMs is significant, rendering the single photo electron (SPE) peaks distinguishable. These peaks can in turn be used to determine the gains of the SiPMs, and consequently stabilize it at a desired value. Fig.~\ref{finger_gain} (left) shows an example of the SPE peaks observed from a charge histogram. To identify each of the SPE peaks, a Gaussian fit is applied, and the distance between the peaks is deduced to determine the gain of the SiPM. A gain versus bias voltage plot for a single calibration run is presented in Fig.~\ref{finger_gain} (right). This calibration data was investigated for a period of 3 months, as a function of the ambient temperature recorded by the detectors. Fig.~\ref{3D} presents the gain of the detectors as a function of the ambient temperature and supplied bias voltage. This relationship is modeled by a 2-dimensional plane of the form 
%$\mathrm{Gain} = a* \mathrm{Voltage} + b* \mathrm{Temp} +c$, 
\begin{equation}
\mathrm{Gain} = a * \mathrm{Voltage} + b * \mathrm{Temp} + c,
\end{equation}
for which the best-fit parameters are: $a$ = -0.36, $b$ = 0.1, $c$ = -241.03. 
This is done for all scintillation detectors, with an objective of stabilizing the gain of the SiPMs which will be implemented with the upcoming version of the data acquisition software for the uDAQ board.

\begin{figure}[ht!]
    % \centering
    \begin{minipage}{0.5\textwidth}
        \centering
        \includegraphics[width=\linewidth]{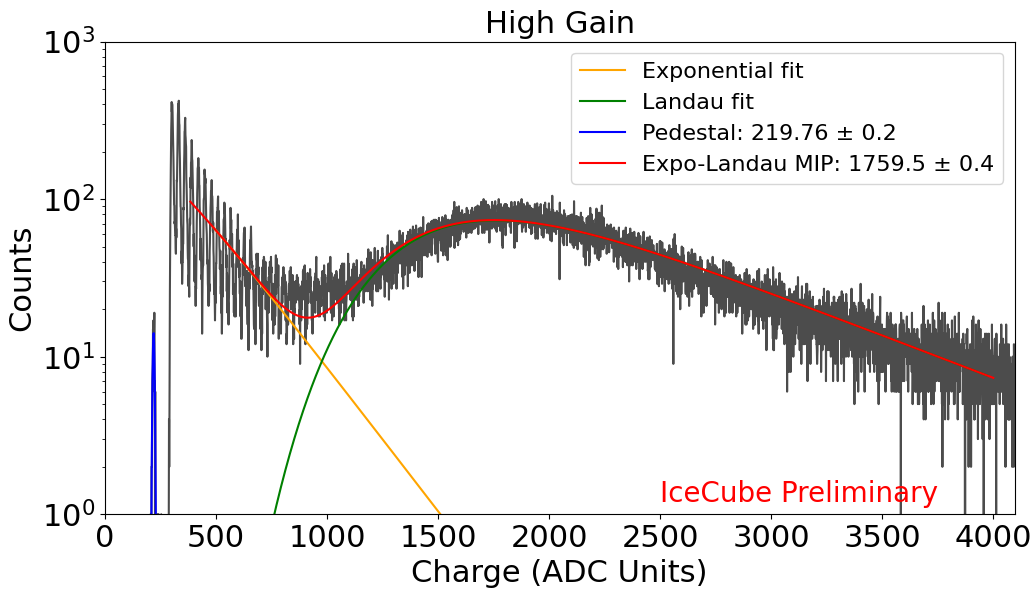}
        \label{fig:prob1_6_2}
    \end{minipage}
    \begin{minipage}{0.5\textwidth}
        \centering
        \includegraphics[width=\linewidth]{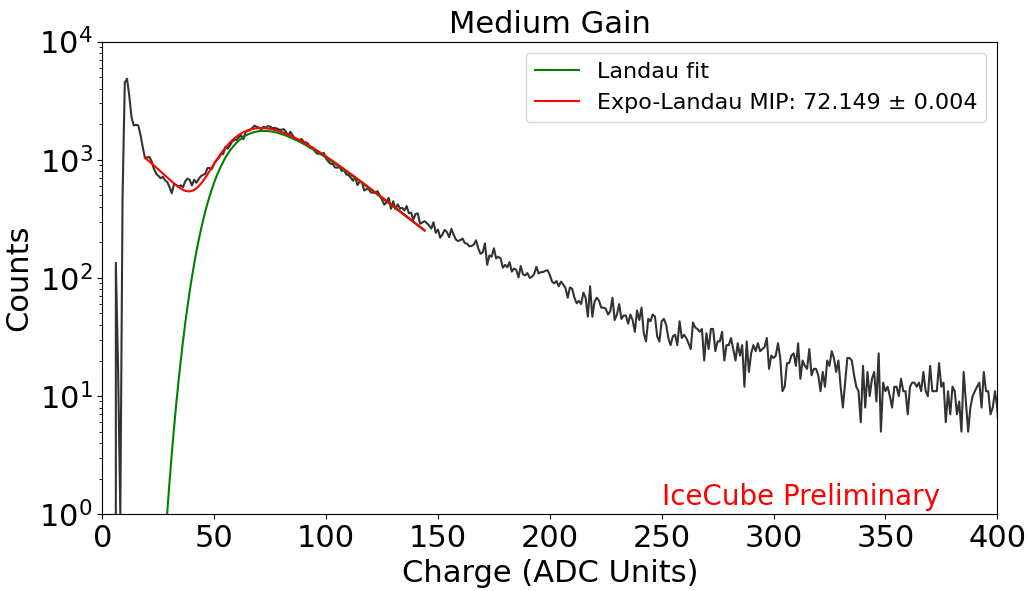}
        \label{fig:prob1_6_1}
    \end{minipage}
    \begin{minipage}{0.5\textwidth}
        \includegraphics[width=\linewidth]{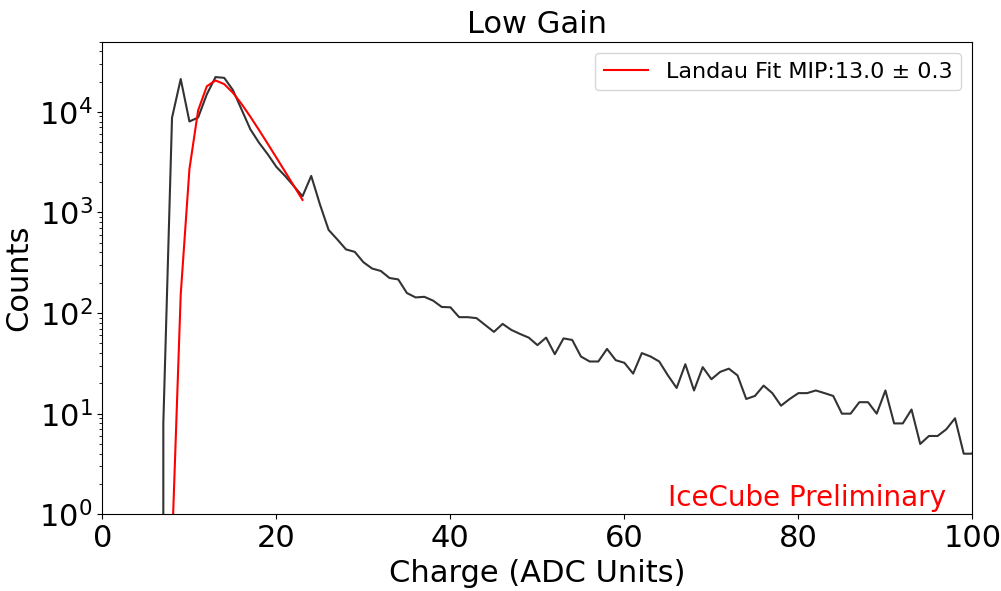}
    \end{minipage}
    \begin{minipage}{0.5\textwidth}
        \centering
        \caption{Example charge histograms in the three uDAQ gains. The high and medium gain are fit with exponential plus Landau function, whereas for the low gain, only the MIP peak is fit with a Landau function. Note the different x-axis ranges.}
        \label{chargespectra}
    \end{minipage}
\end{figure}

The light yield as well as the dynamic range of the detectors can additionally be obtained by the characterisation of the minimally ionizing particle (MIP) peak observed from the charge spectra. Example charge histograms for the three representative uDAQ gains are presented in Fig.~\ref{chargespectra}. The entire charge spectrum can be described by a summation of an exponential and Landau distribution. The exponentially decaying dark noise from the SiPM is combined with the charge deposit from the MIPs, which is expected to follow a Landau distribution. This fitting is applied to the high and medium gain charge histograms. In the lowest gain, the resolution is not sufficient to resolve the two distributions, and therefore the MIP is fit with a Landau distribution only.

The dynamic range is determined using the ratio of the total available channels in a 12 bit ADC and the ADC units per MIP in the lowest gain. For the new detectors, this ratio implies a dynamic range of the order of $\sim$800 MIPs (Fig.~\ref{chargespectra} (bottom): The first filled bin, corresponds to the pedestal in low gain). 
This is an improvement to the prototype station which had a dynamic range of 200~MIPs~\cite{Oehler2022_1000142813}. The light yield depends on the number of photo electrons per MIP, which is for the given example $\sim$55 photo-electrons/MIP. For the specified dynamic range, this renders a total light yield of $\sim$44,000~PEs, which is  within the linear regime for the SiPMs~\cite{Bretz:2017dmu}.

\section{Mass Production}\label{sec1}

\begin{figure}[h]
  \centering
  \includegraphics[scale=0.231]{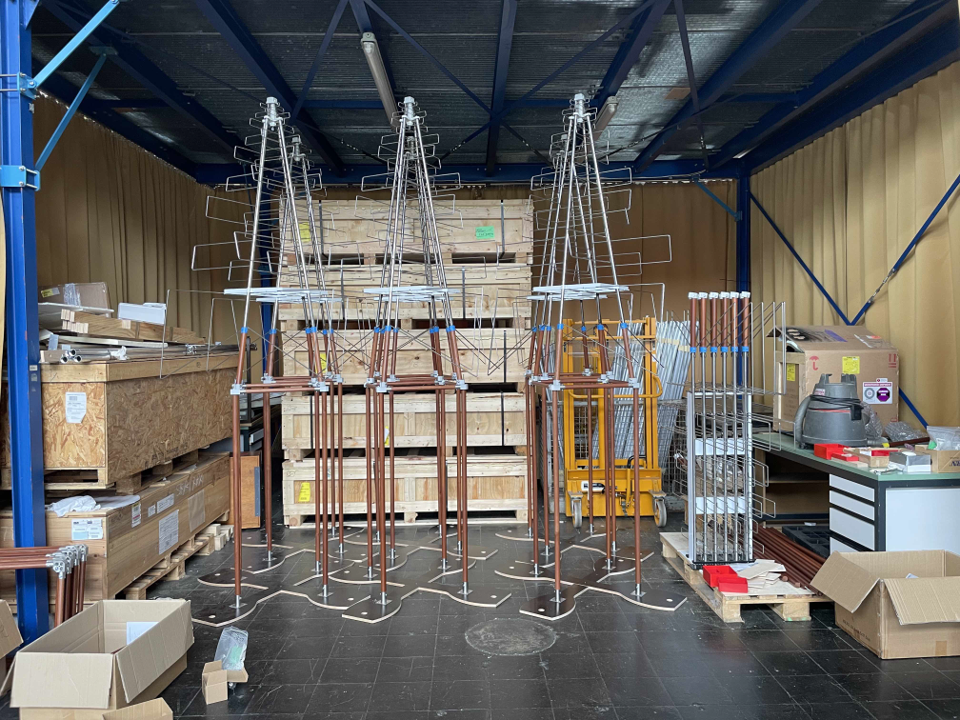}
  \includegraphics[scale=0.125]{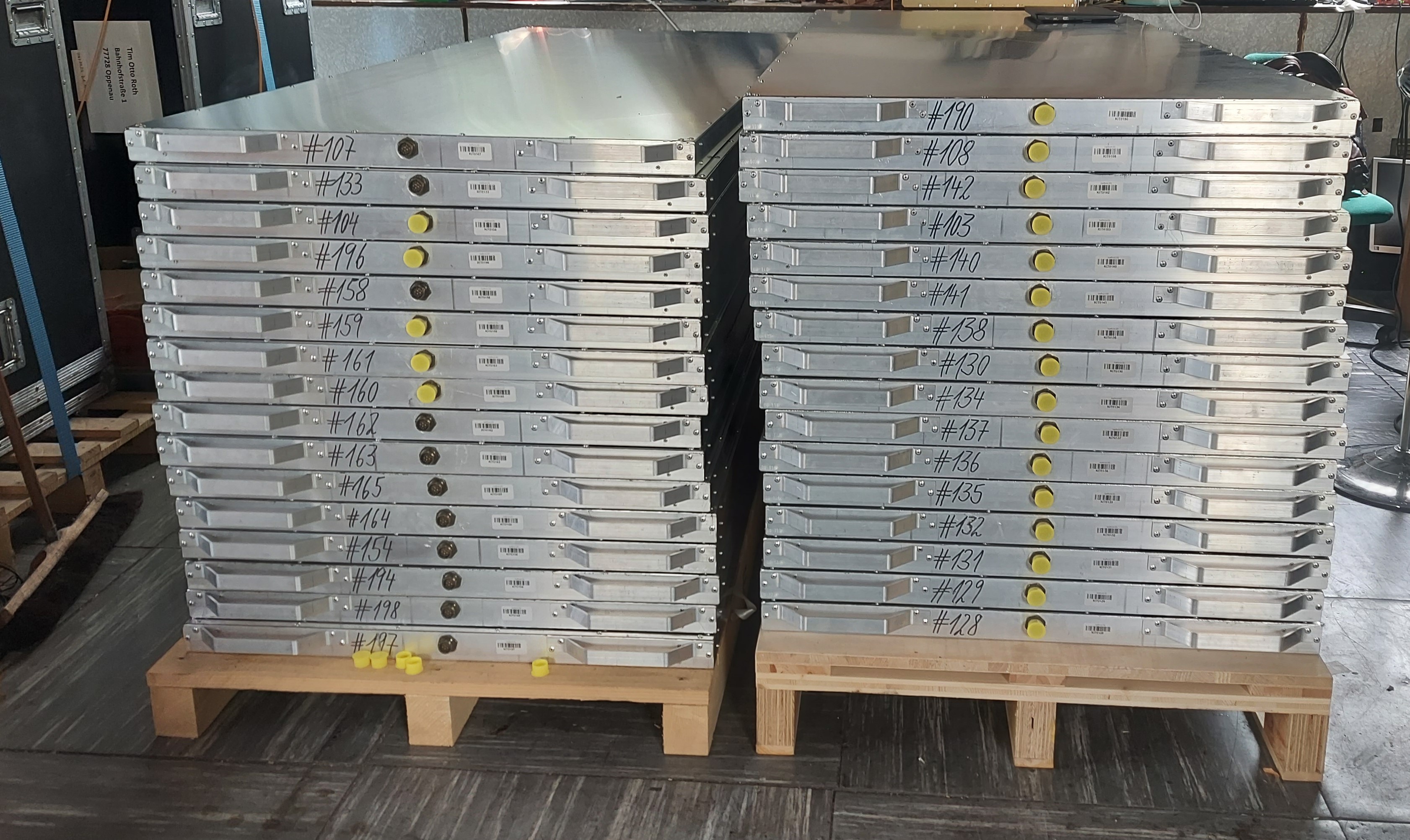}
  \caption{Left: Pre-assembly of the antenna mounts; Right: Fully assembled scintillation detectors ready for calibration tests. }
  \label{mass}
  \vspace*{-0.3cm}
\end{figure}

The deployment of more enhancement stations is proposed in batches of 6 stations. The production of the scintillation and radio detectors for the first phase is complete. Fig.~\ref{mass} shows pictures of the radio antennas and scintillation detectors during the production phase before shipment.

The radio antennas for the 6 stations are pre-assembled, labeled and packed in single station batches. The LNAs are calibrated in a laboratory setup and shipped separately. Since the scintillation detectors are fully assembled before dispatch they have to be calibrated on the production site. Each scintillator is calibrated in a lead shielding setup~\cite{Shefali:2023geo}. The calibration measurements include:

\begin{itemize}
    \item Coincidence measurements to send a trigger to the radio antennas, with scintillator multiplicity set from 1 up to 8 scintillation detectors
    \item Charge histogram measurements in the two available measurement modes on the uDAQ namely, histogram and hitbuffer, explained in ~\cite{Oehler2022_1000142813}
    \item Threshold scans over the entire threshold range of the uDAQ in digital channel units.
\end{itemize}

\begin{figure}[h]
  \centering
  \includegraphics[scale=0.5]{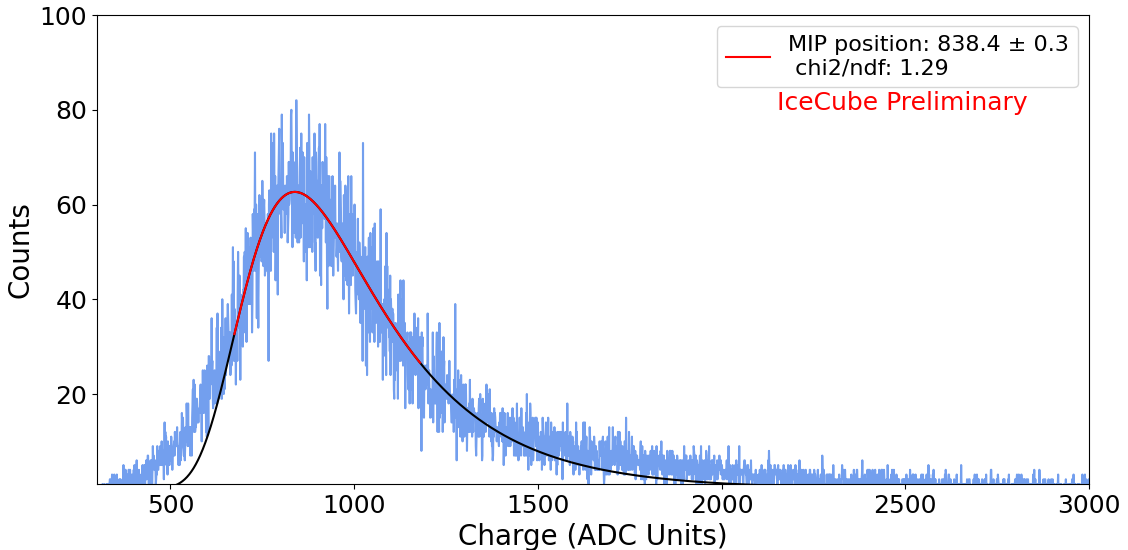}
  \caption{Charge histogram of the filtered events for one station at the production site prior to assembly, fit with a Landau function. The red curve corresponds to the region of the data where the fit is performed.}
  \label{mass_filtered}
  \vspace*{-0.3cm}
\end{figure}

These measurements are used to ensure the functionality of each of the produced scintillation detectors, as well as to calibrate the MIP positions at room temperatures. The room temperature measurements are carried out in a group of 8 scintillators, stacked on top of each other. Due to the high background (originating from natural radioactivity), despite a carefully chosen test setup~\cite{Shefali:2023geo}, the charge histograms from each of the scintillation detectors are filtered to keep entries which correspond to hits that occur within 200~ns in all of the 8 scintillation detectors. Following this, a Landau distribution is fit to the filtered events, to localise the MIP position. Fig.~\ref{mass_filtered} presents the filtered charge histogram for one of the panels in high gain. Following these room-temperature calibration tests, the scintillation detectors are further shipped for low temperature tests prior to the deployment at the South Pole.

\section{R\&D Stations}\label{sec1}

In addition to the enhancement stations, multiple R\&D stations have been produced. One full station each has been deployed at the Pierre Auger Observatory in Malargüe, Argentina and the Telescope Array (TA) in Utah, USA.

\begin{figure}[ht!]
  \centering
  \includegraphics[scale=0.5]{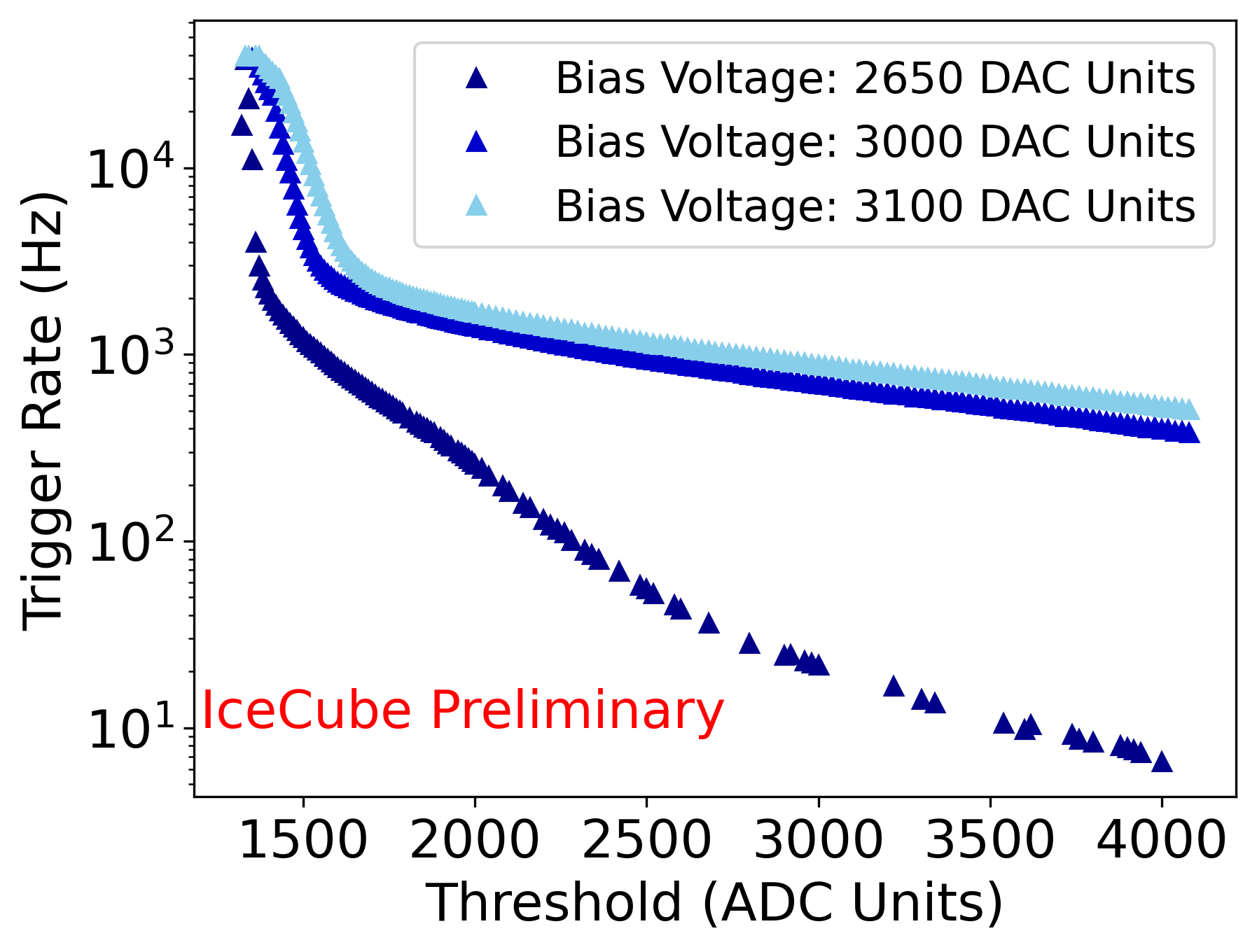}
  \caption{Threshold Scans performed at 3 varied bias voltages.}
  \label{TScan}
  \vspace*{-0.3cm}
\end{figure}

The station at the Pierre Auger Observatory has been operational since December 2022. The main aim of this station is comparing its radio measurements with the Auger Engineering Radio Array (AERA). To achieve that, the optimal trigger rate for the scintillation detectors needs to be determined, since the expected natural radioactivity from the ground, as well as the high ambient temperatures, are unfavorable for the SiPM operation. For this purpose, a threshold scan at various supplied bias voltages to the scintillation detectors is performed. Fig.~\ref{TScan} shows the trigger rate for the 3 different bias voltages in DAC (digital to analog convertor) units. The bias voltage is set to 3000~DAC units to achieve a nominal trigger rate at maximum threshold ($\sim400~\mathrm{Hz}$), while keeping the electronic noise to a minimum. At a higher bias voltage, the electronics are expected to saturate.

The station for the TA has been deployed in the low-energy infill region, with an objective to validate a possibility for cross calibration with the TA scintillation detectors for air shower measurements in the future for a full array~\cite{Goehlke2022_1000153878}. Furthermore, this station will also be used for improvement and adaptation of the central DAQ unit of the enhancement stations for the proposed next generation of the IceCube Observatory, 
IceCube-Gen2~\cite{IceCube-Gen2:2021aek}. This station is not yet in operation.

\section{Summary}\label{sec1}

The prototype station of the Surface Array Enhancement has been taking useful R\&D data at the South Pole.  It has recently been upgraded, and will be the first station in the Surface Array Enhancement planned for the future. The commissioning of this station is in progress. 
Furthermore, promising advancements have been made for the isolation of the radio emission of air shower component from the background with the prototype station data, which will facilitate various cosmic-ray analysis. For the future deployments, 6 full stations have been produced and calibrated. Moreover, multiple developments are presented at major cosmic-ray detector sites,  which will help to further improve the enhancement detectors as well as the DAQ system for IceCube-Gen2.

% Bibtex references:
\bibliographystyle{ICRC}
\bibliography{references}

\clearpage

\section*{Full Author List: IceCube Collaboration}

\scriptsize
\noindent
R. Abbasi$^{17}$,
M. Ackermann$^{63}$,
J. Adams$^{18}$,
S. K. Agarwalla$^{40,\: 64}$,
J. A. Aguilar$^{12}$,
M. Ahlers$^{22}$,
J.M. Alameddine$^{23}$,
N. M. Amin$^{44}$,
K. Andeen$^{42}$,
G. Anton$^{26}$
% ...
\\
\\
% ...
$^{12}$ Universit{\'e} Libre de Bruxelles, Science Faculty CP230, B-1050 Brussels, Belgium \\
% ...
$^{17}$ Department of Physics, Loyola University Chicago, Chicago, IL 60660, USA \\
$^{18}$ Dept. of Physics and Astronomy, University of Canterbury, Private Bag 4800, Christchurch, New Zealand \\
% ...
$^{22}$ Niels Bohr Institute, University of Copenhagen, DK-2100 Copenhagen, Denmark \\
$^{23}$ Dept. of Physics, TU Dortmund University, D-44221 Dortmund, Germany \\
% ...
$^{26}$ Erlangen Centre for Astroparticle Physics, Friedrich-Alexander-Universit{\"a}t Erlangen-N{\"u}rnberg, D-91058 Erlangen, Germany \\
% ...
$^{40}$ Dept. of Physics and Wisconsin IceCube Particle Astrophysics Center, University of Wisconsin{\textendash}Madison, Madison, WI 53706, USA \\
% ...
$^{42}$ Department of Physics, Marquette University, Milwaukee, WI, 53201, USA \\
% ...
$^{44}$ Bartol Research Institute and Dept. of Physics and Astronomy, University of Delaware, Newark, DE 19716, USA \\
% ...
$^{63}$ Deutsches Elektronen-Synchrotron DESY, Platanenallee 6, 15738 Zeuthen, Germany  \\
$^{64}$ Institute of Physics, Sachivalaya Marg, Sainik School Post, Bhubaneswar 751005, India \\
% ...

\subsection*{Acknowledgements}

\noindent
The authors gratefully acknowledge the support from the following agencies and institutions:
USA {\textendash} U.S. National Science Foundation-Office of Polar Programs,
U.S. National Science Foundation-Physics Division,
U.S. National Science Foundation-EPSCoR,
Wisconsin Alumni Research Foundation,
Center for High Throughput Computing (CHTC) at the University of Wisconsin{\textendash}Madison,
Open Science Grid (OSG),
Advanced Cyberinfrastructure Coordination Ecosystem: Services {\&} Support (ACCESS),
Frontera computing project at the Texas Advanced Computing Center,
U.S. Department of Energy-National Energy Research Scientific Computing Center,
Particle astrophysics research computing center at the University of Maryland,
Institute for Cyber-Enabled Research at Michigan State University,
and Astroparticle physics computational facility at Marquette University;
Belgium {\textendash} Funds for Scientific Research (FRS-FNRS and FWO),
FWO Odysseus and Big Science programmes,
and Belgian Federal Science Policy Office (Belspo);
Germany {\textendash} Bundesministerium f{\"u}r Bildung und Forschung (BMBF),
Deutsche Forschungsgemeinschaft (DFG),
Helmholtz Alliance for Astroparticle Physics (HAP),
Initiative and Networking Fund of the Helmholtz Association,
Deutsches Elektronen Synchrotron (DESY),
and High Performance Computing cluster of the RWTH Aachen;
Sweden {\textendash} Swedish Research Council,
Swedish Polar Research Secretariat,
Swedish National Infrastructure for Computing (SNIC),
and Knut and Alice Wallenberg Foundation;
European Union {\textendash} EGI Advanced Computing for research;
Australia {\textendash} Australian Research Council;
Canada {\textendash} Natural Sciences and Engineering Research Council of Canada,
Calcul Qu{\'e}bec, Compute Ontario, Canada Foundation for Innovation, WestGrid, and Compute Canada;
Denmark {\textendash} Villum Fonden, Carlsberg Foundation, and European Commission;
New Zealand {\textendash} Marsden Fund;
Japan {\textendash} Japan Society for Promotion of Science (JSPS)
and Institute for Global Prominent Research (IGPR) of Chiba University;
Korea {\textendash} National Research Foundation of Korea (NRF);
Switzerland {\textendash} Swiss National Science Foundation (SNSF);
United Kingdom {\textendash} Department of Physics, University of Oxford. This project has received funding from the European Research Council (ERC) under the European Union’s Horizon 2020 research and innovation programme (grant agreement No 802729).

\end{document}